\documentclass[aps,prl,twocolumn,superscriptaddress,showkeys]{revtex4-1}

\usepackage{graphicx}
\usepackage{dcolumn}
\usepackage{bm}
\usepackage[dvipsnames]{xcolor}
\usepackage{wasysym}
\usepackage{lipsum}
\usepackage{enumitem}
\usepackage{booktabs} 
\usepackage{gensymb}
\usepackage{xcolor}
\usepackage{amssymb}

\usepackage{blindtext}

\def\code#1{\texttt{#1}}

\def\vecy{\mathbf{y}}

\def\nitro{\mathrm{N}}
\def\oxy{\mathrm{O}}

\makeatletter
\def\p@subsection{}
\makeatother

\begin{document}


\title{Machine-Learning X-ray Absorption Spectra to Quantitative Accuracy}


\author{Matthew R. Carbone}
\affiliation{Department of Chemistry, Columbia University, New York,
New York 10027, USA}

\author{Mehmet Topsakal}
\email{mtopsakal@bnl.gov}
\affiliation{Nuclear Science and Technology Department, Brookhaven National Laboratory, Upton, New York 11973, USA}

\author{Deyu Lu}
\email{dlu@bnl.gov}
\affiliation{Center for Functional Nanomaterials, Brookhaven National Laboratory, Upton, New York 11973, USA}

\author{Shinjae Yoo}
\email{sjyoo@bnl.gov}
\affiliation{Computational Science Initiative, Brookhaven National Laboratory,
Upton, New York 11973, USA}

\date{\today}

\begin{abstract}
Simulations of excited state properties, such as spectral functions, are often computationally expensive and therefore not suitable for high-throughput modeling. As a proof of principle, we demonstrate that graph-based neural networks can be used to predict the x-ray absorption near-edge structure spectra of molecules to quantitative accuracy. Specifically, the predicted spectra reproduce nearly all prominent peaks, with 90\% of the predicted peak locations within 1 eV of the ground truth. Besides its own utility in spectral analysis and structure inference, our method can be combined with structure search algorithms to enable high-throughput spectrum sampling of the vast material configuration space, which opens up new pathways to material design and discovery.
\end{abstract}

\keywords{x-ray absorption spectroscopy, machine learning, neural network, first-principles calculations, molecular structure}

\pacs{}

\maketitle



The last decade has witnessed exploding developments in
artificial intelligence, specifically deep learning applications, in many
areas of our society~\cite{lecun2015review}, including 
image and 
speech recognition,
language translation and drug discovery, just to name a few. In scientific research, deep learning methods allow
researchers to establish rigorous, highly non-linear relations in high-dimensional data. This enormous potential has been
demonstrated in, e.g., solid state physic and materials science~\cite{reyes2019machine,schmidt2019recent}, including the
prediction of molecular~\cite{Rupp2012,rupp2015machine} and crystal~\cite{XIE2018} properties, infrared~\cite{gastegger2017machine} and optical excitations~\cite{ye2019neural}, phase transitions~\cite{vargas2018} and
topological ordering~\cite{rodriguez2019topo} in model systems, \emph{in silico} materials
design~\cite{sanchez2018inverse} and force field development~\cite{Behler2007,zhang2018deep}.

One high-impact area of machine learning (ML) applications 
is predicting material properties. By leveraging large amounts of labeled data consisting of feature-target pairs, ML
models, such as deep neural networks, are trained to map features to targets. The ML parameters are
optimized by minimizing an objective loss criterion, and yields a locally optimal interpolating
function~\cite{introMLtextbook}. Trained ML models can make accurate predictions on unknown materials almost instantaneously, giving this approach a huge advantage in terms of fidelity and efficiency in sampling the vast materials
space as compared to experiment and conventional simulation methods. So far, existing ML predictions mostly focus on
simple quantities, such as the total energy, fundamental band gap and forces; it remains unclear whether ML models
can predict complex quantities, such as spectral functions of real materials, with high accuracy. Establishing such
capability is in fact essential to both the physical understanding of fundamental processes and design of new
materials. In this study, we demonstrate that ML models can predict x-ray absorption spectra of molecules with
quantitative accuracy, capturing key spectral features, such as locations and intensities of
prominent peaks.

X-ray absorption spectroscopy (XAS) is a robust, element-specific characterization technique widely used to
probe the structural and electronic properties of materials~\cite{ankudinov2002sensitivity}. 
It measures the intensity loss of incident light through the sample
caused by core electron excitations to unoccupied states~\cite{rehr2000rmp}.
In particular, the x-ray absorption near edge structure
(XANES) encodes key information about the local
chemical environment (LCE), e.g. the charge state, coordination number and local symmetry, of the absorbing
sites~\cite{kuzmin2014exafs,rehr2000rmp,rehr2009ab}. Consequently, XANES is a premier method for studying
structural changes, charge transfer, and charge and magnetic ordering in
condensed matter physics, chemistry and materials science. 

To interpret XANES spectra, two classes of problems need to be addressed. In a \emph{forward} problem, one
simulates XANES spectra from given atomic arrangements using electronic structure
theory~\cite{rehr2000rmp, Mathieu2002, Prendergast2006, deGroot2008,Chen2010, ocean2011,gulans2014exciting}. In an \emph{inverse} problem, one infers key LCE
characteristics from XANES spectra~\cite{Timoshenko2017,Timoshenko2018,Carbone2019}. 
While the solution of the forward problem is limited by the accuracy of the theory and computational expense, it is generally more complicated to solve the inverse problem, which often suffers from a lack of
information and can be ill-posed~\cite{rehr2005bayes}. Standard approaches typically rely on either empirical fingerprints from experimental references of known crystal structures or verifying hypothetical models using forward
simulation~\cite{farges1997prb,Farges2001}.

When using these standard approaches, major challenges arise from material
complexity associated with chemical composition (e.g., alloys and doped materials) and structure (e.g., surfaces, interfaces and defects), which makes it impractical to find corresponding reference systems
from experiment and incurs a high computational cost of simulating a large number of possible
configurations, with hundreds or even thousands of atoms in a single unit cell. Furthermore, emerging
high-throughput XANES capabilities~\cite{meirer2018spatial} poses new challenges for fast, even on-the-fly, solutions of
the inverse problem to provide time-resolved materials characteristics for \emph{in situ} and \emph{operando}
studies. As a result, a highly accurate, high-throughput XANES simulation method could play a crucial role in tackling
both forward and inverse problems, as it provides a practical means to navigate the material space in order to unravel
the structure-spectrum relationship. When combined with high-throughput structure sampling methods, ML-based
XANES models can be used for the fast screening of relevant structures.

Recently, multiple efforts have been made to incorporate data science tools in x-ray spectroscopy.
Exemplary studies include database infrastructure development (e.g. the computational XANES database in the
Materials Project~\cite{Jain2013,Ong2013,Ong2015,Mathew2018}), building computational spectral
fingerprints~\cite{yan2019ultrathin}, screening local structural motifs~\cite{trejo2019elucidating}, predicting LCE
attributes in nano clusters~\cite{Timoshenko2017} and crystals~\cite{Timoshenko2018, Carbone2019} from XANES spectra using
ML models. However, predicting XANES spectra directly from molecular structures using ML models has, to
the best of our knowledge, not yet been attempted.

As a proof-of-concept, we show that a graph-based deep learning architecture, a message passing neural
network (MPNN)~\cite{MPNN_theory}, can predict XANES spectra of molecules from their molecular structures to
quantitative accuracy. Our training sets consist of O and N K-edge XANES spectra (simulated using the~\code{FEFF9}
code~\cite{Rehr2010}) of molecules in the QM9 molecular database~\cite{qm9-2}, which contains
$\sim$~134k small molecules with up to nine heavy atoms (C, N, O and F) each. The
structures were optimized using density functional theory with the same functional and numerical convergence criteria.
This procedure, together with the atom-restriction of the QM9 database, ensures a consistent level of
complexity from which a ML database can be constructed and tested. Although our model is trained on
computationally inexpensive \code{FEFF} data, it is straightforward to generalize this method to XANES spectra simulated at different levels of theory.

The MPNN inputs (feature space) are derived from a subset of
molecular structures in the QM9 database, henceforth referred to as the
\emph{molecular structure space}, $\mathcal{M}$. Two separate databases are constructed by choosing molecules containing at least one O ($\mathcal{M}_\oxy$, $n_\oxy \approx 113$k) or at least one N atom
($\mathcal{M}_\nitro$, $n_\nitro \approx 81$k) each; note that $\mathcal{M}_\oxy \cap
\mathcal{M}_\nitro \neq \emptyset,$ as many molecules contain both O and N atoms. 
The molecular geometry and chemical properties of each molecule are mapped to a graph
($\mathcal{M}_A \rightarrow \mathcal{G}_A$, $A \in \{\oxy, \nitro\}$) by associating atoms with
graph nodes and bonds with graph edges. Following Ref.~\onlinecite{MPNN_theory}, 
each $g_i \in \mathcal{G}_A$ ($i$ the index of the molecule) consists of
an adjacency matrix that completely characterizes the graph connectivity, a list of atom
features (absorber, atom type, donor/acceptor status, and hybridization), and a list of bond
features (bond type and length). A new feature, ``absorber", is introduced to distinguish the absorbing sites from the rest of the nodes. Each graph-embedded molecule in $\mathcal{G}_A$
corresponds to a K-edge XANES spectrum in the \emph{spectrum} or \emph{target space}, $S_A \in
\mathbb{R}^{n_A \times 80}$, which is the average of the site-specific spectra of all absorbing
atoms, $A,$ in that molecule, spline interpolated onto a grid of 80 discretized points and scaled to a maximum intensity of 1.
For each database $\mathcal{D}_A = (\mathcal{G}_A, S_A),$ the data is partitioned into training,
validation and testing splits.
The latter two contain 500 data points each, with the remainder used for training. The MPNN model is optimized using the
mean absolute error (MAE) loss function between the prediction $\hat\vecy_i = \mathrm{MPNN}(g_i)$ and ground truth
$\vecy_i \in S_A$ spectra. During training, the MPNN learns effective atomic properties, encoded in hidden state
vectors at every atom, and passes information through bonds via learned messages. The output computed from the hidden state vectors is the XANES spectrum discretized on the energy grid as a length-80 vector. Additional details regarding the graph embedding
procedure, general implementation~\cite{pytorch,networkx,gru} and MPNN operation can
be found in Ref.~\onlinecite{MPNN_theory} and in the supporting information (SI)~\cite{supplemental}.

\begin{figure}[!tbh]
\includegraphics[width=1.0\columnwidth]{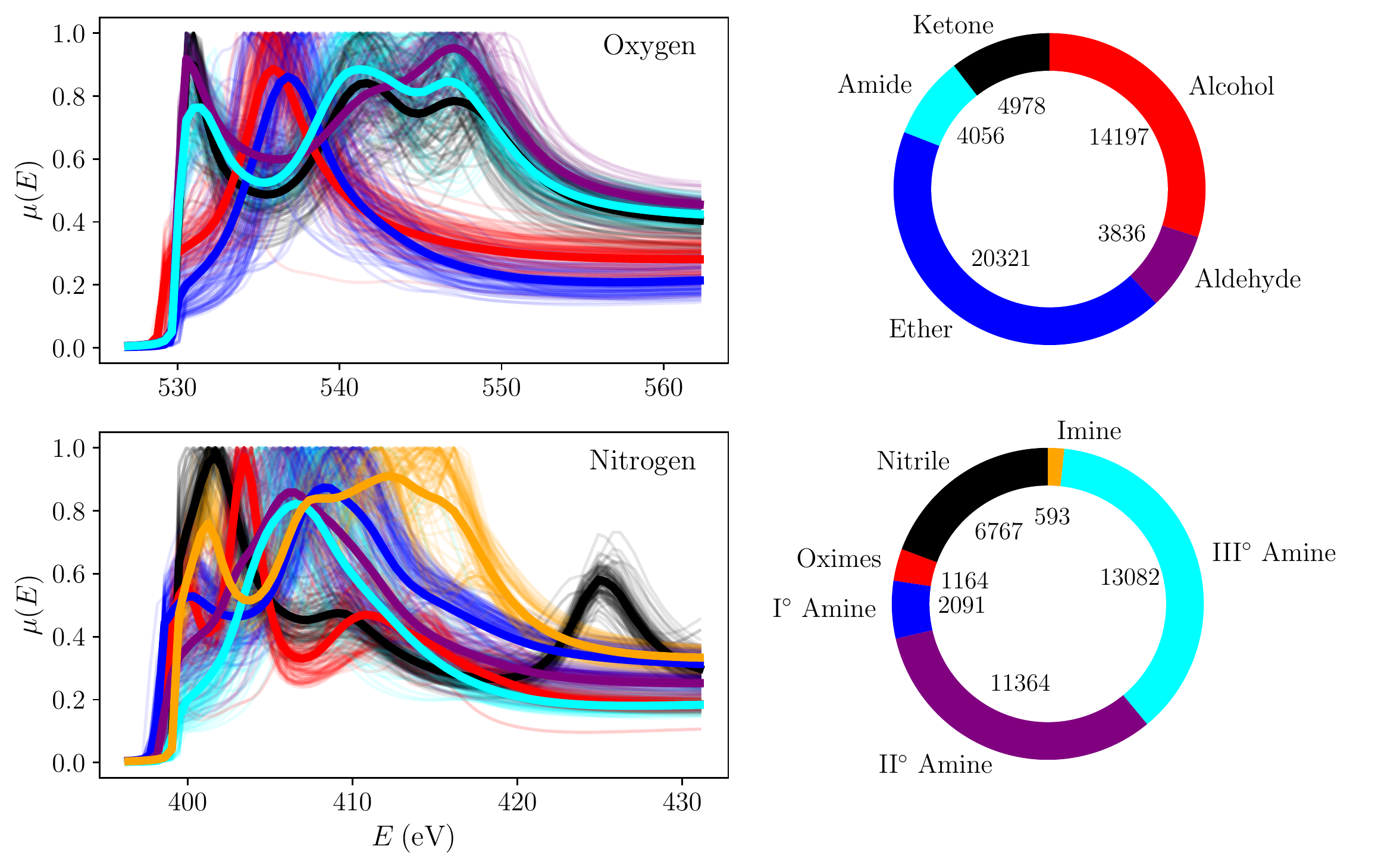}
\caption{\label{fig:legend}
Left: 100 oxygen (top) and nitrogen (bottom) random sample spectra from each functional group in $S_A'$; the averages over all spectra in each functional group are shown in bold. Right: the distribution of functional groups in $\mathcal{D}_A'$.}
\end{figure}

Prior to the training, we systematically examine the distribution of the data. Following common chemical intuition,
the data are labeled according to the functional group that the absorbing atom belongs to. In order to efficiently
deconvolute contributions from different functional groups, we only present results on
molecules with a \emph{single} absorbing atom each; this subset is denoted as
$\mathcal{D}_A' = (\mathcal{G}_A', S_A') \subset \mathcal{D}_A$, and the distribution of common
functional groups in $\mathcal{D}_A'$ are shown in Fig.~\ref{fig:legend}, where the most abundant compounds are ethers and alcohols in $\mathcal{D}_\oxy'$, and tertiary (III$^\circ$) and secondary (II$^\circ$) amines in $\mathcal{D}_\nitro'$.
From averaged spectra (bold lines) in Fig.~\ref{fig:legend}, distinct spectral contrast (e.g., number of prominent peaks, peak
locations and heights) can be identified between different functional groups. In fact, several trends in the
\code{FEFF} spectra qualitatively agree with experiment, such as the sharp pre-edge present in ketones (black)
but absent in alcohols (red)~\cite{kim2011characterization}, and the general two-peak feature of primary (I$^\circ$) amines (blue)~\cite{plekan2007x}.

\begin{figure}[!htb]
\includegraphics[width=1.\columnwidth]{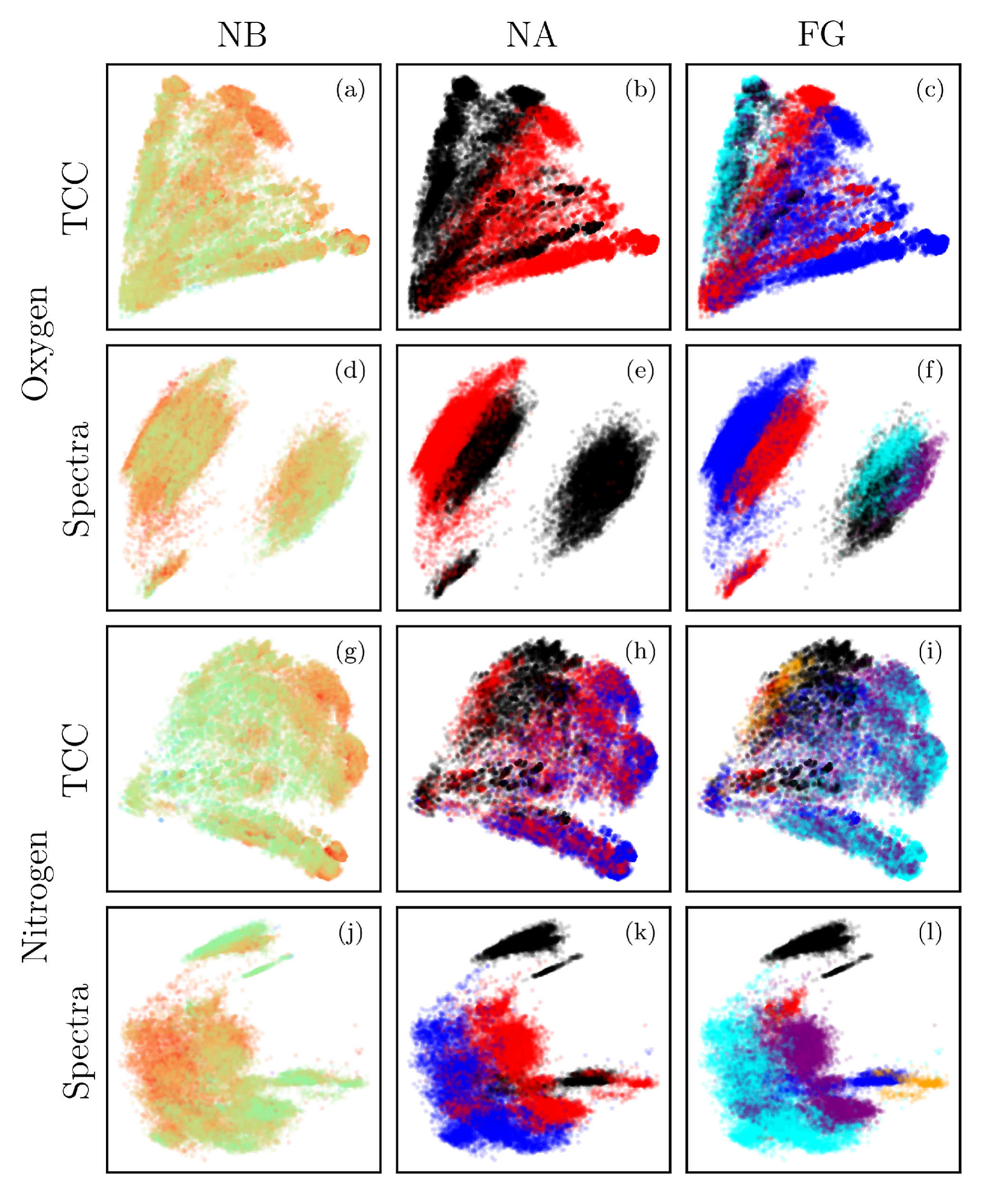}
\caption{\label{fig:pca}
PCA plots for both the TCC and spectra proxies for the molecules in $\mathcal{D}'_A$ labeled by NB, NA and FG. The total number of non-hydrogenic bonds (NB, top) range from 1 (violet) to 13 (red). The total number of atoms bonded to the absorbing atom (NA, center) takes on one of three values: 1, 2 or 3 (black, red and blue, respectively). The color legends for the functional group of the absorbing atom (FG, bottom) are
the same as in Fig.~\ref{fig:legend}.}
\end{figure}

Although XANES is known as a local probe that is sensitive to the LCE of absorbing atoms, a
systematic study of the degree of such correlation on a large database has not yet been performed. To investigate this structure-spectrum correlation, we perform principal component analysis (PCA) \cite{pca} on both the
features and targets in $\mathcal{D}_A$, and visually examine the clustering patterns after the data
in $\mathcal{D}_A'$ is labeled by different chemical descriptors.
To provide a baseline, we consider the total number of
non-hydrogenic bonds in the molecule (NB), which is a generic, global property, supposedly having little relevance to the
XANES spectra. Next we consider two LCE attributes: the total number of atoms bonded to the absorbing atom (NA) and the
functional group of the absorbing atom (FG). While spectra on a discrete grid can
be processed directly, molecular structures, with different number of
atoms and connectivity, need to be pre-processed into a common numerical representation before
clustering. Thus, the molecular fingerprint of each molecule in $\mathcal{M}_A$ is calculated from its SMILES code using the \code{RDKit} library~\cite{rdkit}. Then
an arbitrarily large subset of $10^4$ molecules, $\widetilde{\mathcal{M}}_A \subset \mathcal{M}_A$, is randomly selected
to construct a molecular similarity matrix of Tanimoto correlation coefficients (TCCs)~\cite{tanimoto1958elementary},
$T_A \in [0, 1]^{N_A \times 10^4}$, from the molecular fingerprints such that 
$T_{A,ij} = \mathrm{TCC}(m_i, m_j)$, where $m_i \in \mathcal{M}_A$ and $m_j \in \widetilde{\mathcal{M}}_A$.
$\mathrm{TCC}(m_i, m_i) = 1$ defines perfect similarity. The
$T_A$ matrix therefore provides a uniform measure of structural similarity of every molecule in $\mathcal{M}_A$ to each
one of the $10^4$ references, serving as a memory-efficient proxy to $\mathcal{M}_A$.

Results of the PCA dimensionality reduction are presented for both data sets
and all three descriptor labels (NB, NA and FG) in Fig.~\ref{fig:pca}. Specifically, after PCA is
performed on unlabeled data, the data are colored in by their respective labels.
While some degree of structure is manifest in NB, it is clear that the overall clustering is much inferior to both NA and FG, confirming that NB is largely irrelevant to XANES. On the other hand, both NA and FG exhibit significant clustering, with the latter, as expected, slightly more resolved; while NA can only distinguish up to 2 (3) bonds in the O (N) data sets, FGs reveal more structural details of the LCE, and encode more precise information, such as atom and bond types. For NA and FG, clustering in the TCC-space is more difficult to resolve, as it is only a course-grained description of the molecule, missing detailed information about, e.g., molecular geometry, which will be captured by the MPNN. Despite this, visual inspection reveals significant structure, such as in Fig.~\ref{fig:pca}(c), where alcohols (red), ethers (blue) and amides (cyan) appear well-separated.

Spectra PCA of FG in Figs.~\ref{fig:pca}(f) and \ref{fig:pca}(l) can also be directly correlated with the sample spectra in Fig.~\ref{fig:legend}. For instance, the shift in the main peak position between ketones/aldehydes/amides (black/purple/cyan) and alcohols/ethers (red/blue) in $S_\oxy'$ reflects the impact of a double versus a single bond on
the XANES spectra. As a result, groups of these structurally different compounds are well-separated in the spectra PCA as shown in Fig.~\ref{fig:pca}(f); even compounds with moderate spectral contrast, e.g., between alcohols (red) and ethers (blue), are well-separated. Similar trends are observed in $S_\nitro',$ where, e.g.,
nitrile groups (black) show a distinct feature around 425 eV, which clearly distinguishes itself from the other FGs, and, likely because of that, one observes a distinct black cluster in Fig.~\ref{fig:pca}(l).

\begin{figure}[!htb]
\includegraphics[width=1.0\columnwidth]{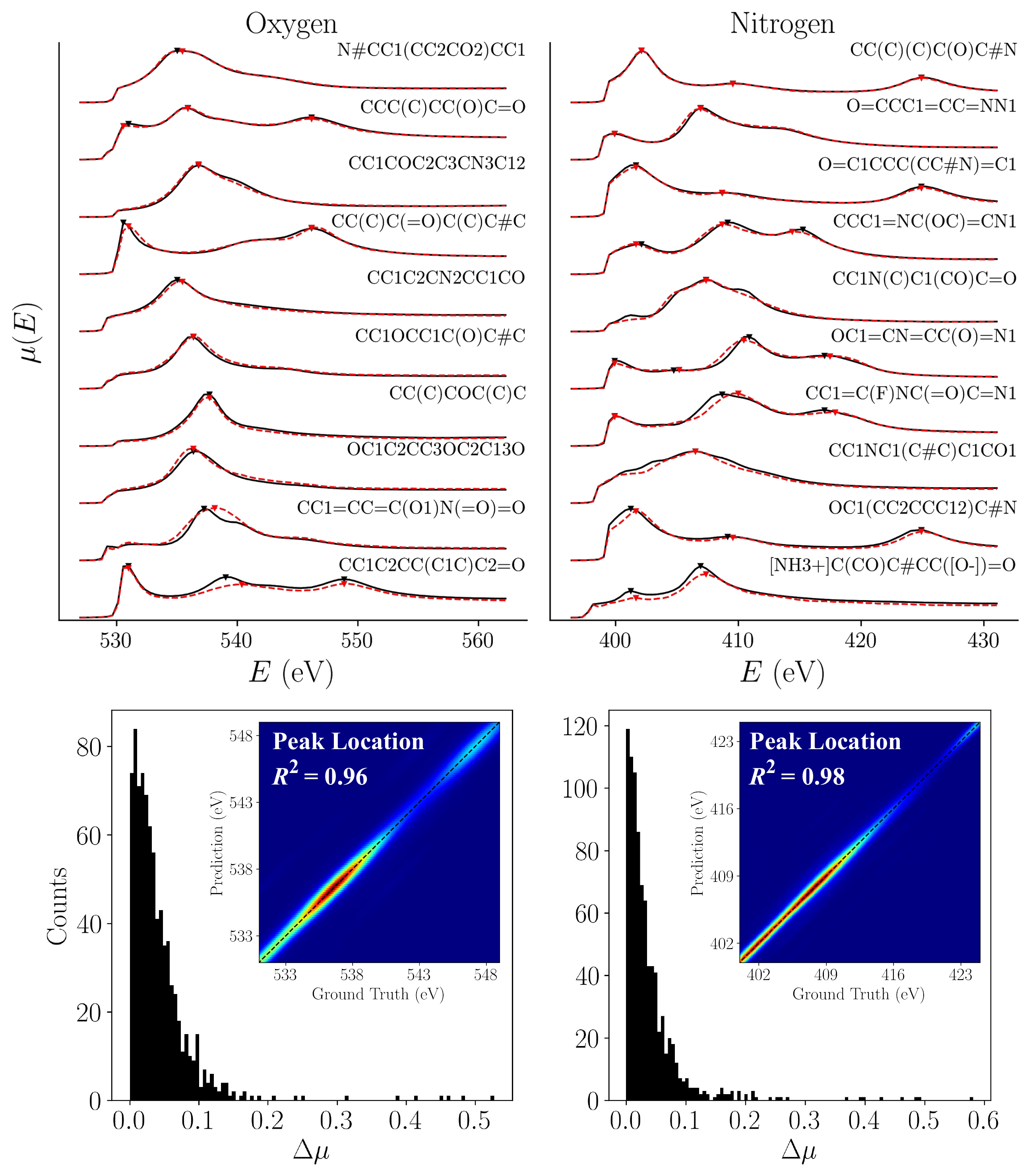}
\caption{\label{fig:waterfall}
Performance metrics for the MPNN evaluated on the $\mathcal{D}_A$ testing sets. Top: waterfall plots  of sample spectra
(labeled by their SMILES codes) of ground truth (black) and predictions (dashed red), where prominent peaks (see text)
are indicated by triangles. One randomly selected sample from every decile is sorted by MAE
(first: best; last: worst). Bottom: distribution of the absolute error of predicted peak heights, $\Delta \mu$; insets show the comparison between the prediction and ground truth in peak locations.}
\end{figure}

The PCA suggests that the FG is a key descriptor of XANES. As the MPNN can fully capture the distinction of
FGs through node features, edge features and the connectivity matrix, we expect that an MPNN can learn XANES
spectra of molecules effectively. Randomly selected testing set results from the trained MPNN for both
$\mathcal{D}_\oxy$ and $\mathcal{D}_\nitro$ are presented in Fig.~\ref{fig:waterfall} and ordered according to
MAE, with the best decile at the top and worst decile at the bottom. It is worth noting that MPNN predictions not
only reproduce the overall shape of the spectra, but, more importantly, predict peak locations and heights accurately. In the best decile, the MPNN predictions and ground truth spectra are nearly indistinguishable. Even
in the worst decile, the main spectral features (e.g. three peaks between 530 and 550 eV in the oxygen K-edge and
two peaks between 400 and 410 eV in nitrogen K-edge) are correctly reproduced with satisfactory relative peak
heights.

As shown in Table I, the MAE of the prediction is 0.023 (0.024) for the oxygen (nitrogen) test set, which is an order of magnitude smaller than the spectral variation defined by the mean absolute deviation of the oxygen (0.131) and nitrogen (0.123) test sets.
To provide an additional quantification of the model's accuracy, we select prominent peaks, defined by those with height above half the maximum height of the spectrum and separated by a minimum 12 grid points ($\approx 6$ eV) in energy. We find that the number of prominent peaks in 95\% (90\%) of predicted spectra correspond with that of the ground truth for the oxygen (nitrogen) testing set. Peak
locations and heights are predicted with average absolute difference of
$\overline{\Delta E} = 0.49$ (0.48) eV and $\overline{\Delta \mu} = 0.045$ (0.041), respectively
(see Table~\ref{tab:performance}). The predicted peak heights display a very narrow distribution
around $\Delta \mu = 0$, as the total population in the tail region with $\Delta \mu > 0.1$ is only 7\%
(see Fig.~\ref{fig:waterfall}, bottom).
As shown at the insets, the vast majority ($\sim$90\%) of the predicted peak locations
fall within $\pm$ 1 eV of the ground truth, with the coefficient of determination, $R^2 \geq 0.96$. The exceptional accuracy of the MPNN model results on predicting both peak
location and intensity underscores its predictive power and its ability to capture essential spectral
features.

It is also important to understand the robustness of the network for practical applications; specifically, we examine how
distorting or removing certain features impacts the model performance. To do so, we train separate MPNN models
using ``contaminated" features, where either (1) the bond length is randomized (RBL), or (2) the atom type is
randomly chosen, and all other atomic features are removed (RAF). In addition, we investigate the impact of the locality
in the MPNN prediction of XANES spectra of molecular systems. By default, the MPNN operates on the graph-embedding of the
whole molecule, referred to as the core results. However, the significance of the FG as a sound proxy for the XANES
spectra (see Fig.~\ref{fig:pca}) suggests that local properties, such as the LCE, play a dominant role. Therefore, spatially truncated graphs are likely to be sufficient to predict the XANES spectra of molecules accurately. To quantify this effect, we impose different distance cutoffs ($d_\mathrm{c}$) from $2$ to $6$~\AA~around the absorbing atoms, and train separate ML models using spatially truncated graphs.

\begin{table}[!htb]
\caption{\label{tab:performance} Performance metrics based on the
MAE of the spectra, $\overline{\Delta E}$ and $\overline{\Delta \mu}.$}
\begin{ruledtabular}
\begin{tabular}{lllll}
$A$ & Data & MAE & $\overline{\Delta E}$ (eV) & $\overline{\Delta \mu}$ \\
\colrule
O&\textrm{Core} & 0.023(1) & 0.52(4) & 0.044(2) \\
&\textrm{RBL} & 0.031(1) & 0.55(3) & 0.051(2) \\
&\textrm{RAF} & 0.041(2) & 0.63(3) & 0.068(3) \\
&$d_\mathrm{c}=4~\text{\AA}$ & 0.023(1) & 0.45(3) & 0.040(2) \\
&$d_\mathrm{c}=3~\text{\AA}$ & 0.025(1) & 0.48(3) & 0.040(2) \\
&$d_\mathrm{c}=2~\text{\AA}$ & 0.095(4) & 0.80(4) & 0.179(6) \\
\colrule
N&\textrm{Core} & 0.024(1) & 0.47(3) & 0.042(2) \\
&\textrm{RBL} & 0.029(1) & 0.57(3) & 0.049(2)  \\
&\textrm{RAF} & 0.045(2) & 0.70(4) & 0.084(3)\\
&$d_\mathrm{c}=4~\text{\AA}$ & 0.023(1) & 0.43(3) & 0.039(2)\\
&$d_\mathrm{c}=3~\text{\AA}$ & 0.027(2) & 0.47(3) & 0.046(3)\\
&$d_\mathrm{c}=2~\text{\AA}$ & 0.056(4) & 0.66(4) & 0.099(5)
\end{tabular}
\end{ruledtabular}
\end{table}

Independent MPNN models were trained and tested on each database corresponding to either RBL, RAF and different
$d_\mathrm{c}$ values. As shown in Table~\ref{tab:performance}, 
randomizing the bond length feature does not affect
the performance of MPNN, as $\overline{\Delta E}$ and $\overline{\Delta \mu}$ in RBL only worsen slightly. Atomic features have a larger
impact than the bond length, as $\overline{\Delta E}$ and $\overline{\Delta \mu}$ in RAF have a sizable increase from 0.52 (0.47) to 0.63 (0.70) eV and from 0.044 (0.042) to 0.068 (0.084) in $\mathcal{D}_\oxy$ ($\mathcal{D}_\nitro$). In fact, despite the seemingly large increase, $\overline{\Delta E}$ is still well below 1 eV, i.e., falling within 1-2 grid points, resulting in only a marginal impact on its practical utility. Percentage-wise, the change in $\overline{\Delta \mu}$ is comparable to $\overline{\Delta E}$ for RAF. If we consider relative peak intensity instead of absolute peak intensity as measured by $\Delta \mu$, this difference becomes less significant.

The analysis above leads to a seemingly counter-intuitive conclusion that key XANES features can be obtained with
little knowledge about the atomic features and bond length, 
especially if one considers the importance to know which atoms are the absorption sites.
It turns out that this is not entirely surprising, since it has been shown
that the distinct chemical information of atoms can be extracted by ML techniques from merely the chemical formula
of the compound~\cite{zhou2018learning}, i.e., specific atomic information can be learned through its environment.
In this case, the connectivity matrix likely compensates for a lack of atom-specific information, and supplies
enough knowledge about the LCE to make accurate predictions. As for the effect of the locality, we found that the
results are statistically indistinguishable from the core results when $d_\mathrm{c} \ge 4$~\AA, and breaks down at
$d_\mathrm{c} \approx 2$~\AA,~indicating that the MPNN architecture requires at least the first two coordination
shells to make accurate predictions. 



In summary, we show that the functional group carries statistically significant information about the XANES
spectra of molecules, and that by using a graph-based deep learning architecture, molecular XANES spectra can be
effectively learned and predicted to quantitative accuracy. 
With proper generalization, this method can be used to provide a general purpose, high-throughput capability for predicting spectral information, which may not be limited to XANES, of a broad range of materials including molecules, crystals and interfaces.

M.R.C. acknowledges support from the U.S. Department of Energy through the Computational Sciences Graduate Fellowship (DOE CSGF) under grant number: DE-FG02-97ER25308.  This research used resources of the Center for Functional Nanomaterials, which
is a U.S. DOE Office of Science Facility, and the Scientific Data
and Computing Center, a component of the Computational
Science Initiative, at Brookhaven National Laboratory  under Contract No. DE-SC0012704. The authors acknowledge fruitful discussions with Mark Hybertsen, Xiaohui Qu, Xiaochuan Ge and Martin Simonovsky.

\providecommand{\noopsort}[1]{}\providecommand{\singleletter}[1]{#1}%

\end{document}


\title{Machine-Learning X-ray Absorption Spectra to Quantitative Accuracy \\ Supplemental Information}

\author{Matthew R. Carbone}
\affiliation{Department of Chemistry, Columbia University, New York,
New York 10027, USA}

\author{Mehmet Topsakal}
\email{mtopsakal@bnl.gov}
\affiliation{Nuclear Science and Technology Department, Brookhaven National Laboratory, Upton, New York 11973, USA}

\author{Deyu Lu}
\email{dlu@bnl.gov}
\affiliation{Center for Functional Nanomaterials, Brookhaven National Laboratory, Upton, New York 11973, USA}

\author{Shinjae Yoo}
\email{sjyoo@bnl.gov}
\affiliation{Computational Science Initiative, Brookhaven National Laboratory,
Upton, New York 11973, USA}

\date{\today}

\pacs{}

\maketitle

\section{Feature Generation of the Graph}

\begin{figure*}[!htb]
\includegraphics[width=\columnwidth]{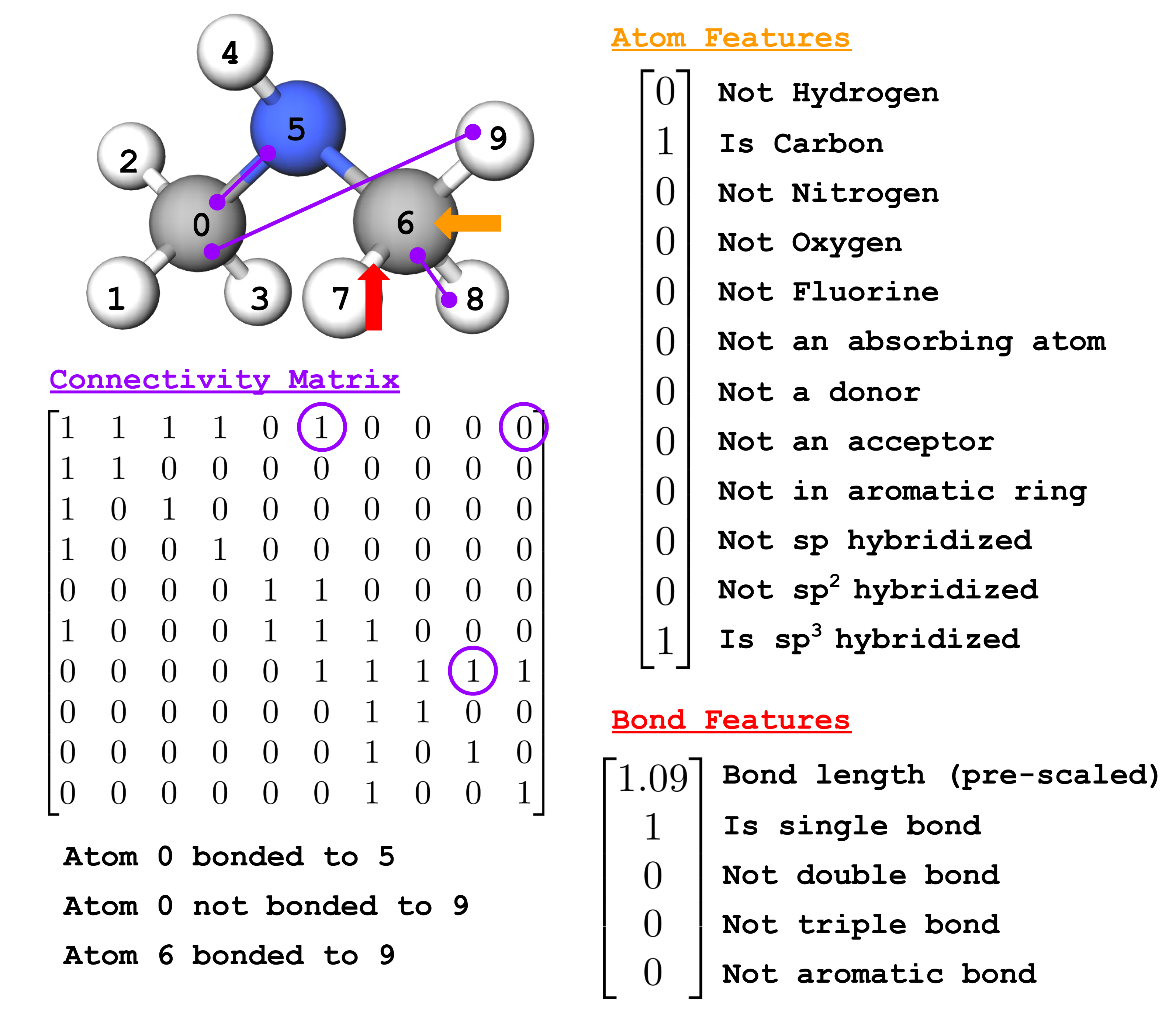}
\caption{\label{fig:molecule_features} Molecular features of an exemplary molecule as processed by the MPNN.
The connectivity matrix (zero-indexed) of the molecule is shown in addition to the features of
atom 6 and bonds $6 - 7$. Specific connectivity is highlighted with purple lines on the molecule, and the
corresponding entries circled in the connectivity matrix.}
\end{figure*}

Starting with an \code{.xyz} geometry file, we performed the following steps to extract the features
described in Table~\ref{tab:feature_list}. To aid the reader in understanding how the feature extraction is
performed, we also present an example in Fig.~\ref{fig:molecule_features}.
Note this can be thought of as performing the mapping
from the molecule-space $\mathcal{M}$ to the graph-space $\mathcal{G}.$
\begin{enumerate}
    \item Using the geometry SMILES code, an \code{RDKit} \cite{rdkit} molecule class was used to
    determine atom donor and acceptor status, aromaticity, atomic number, and hybridization. An extra
    feature (``Absorber") was also included, and indexes whether or not the atom is of type $A$ and
    contributes to computing the target spectra.
    \item A \code{NetworkX} \cite{networkx} graph is constructed from the \code{RDKit} molecule
    class. Edges, along with their edge-types (single, double or triple bonded, or part of an
    aromatic ring) were only initialized between bonded atoms. Bond distances were computed directly
    from the QM9 geometry files.
    \item Adjacency matrices $Q_i$ were initialized for every molecule $m_i$. 
    Matrix elements $Q_{i,jk} = 1$ if
    atom $j$ is bonded to atom $k,$ and $=0$ otherwise.
\end{enumerate}

\begin{table}[!htb]
\caption{\label{tab:feature_list}%
Summary of the input features. The first six denote node (atom) features $\mathbf{V}_i$, 
the next two are edge (bond) features $\mathbf{E}_i$ and last is the adjacency matrix $Q_i$, 
which encodes all atomic connectivity. *Hydrogen atoms are initialized with a one-hot vector of all
zeros. This serves two purposes: both indicating a lack of hybridization and forcing the MPNN to weight
hydrogen atoms slightly less compared to the heavy atoms in the molecule.
**Each bond type can be a single, double, triple or aromatic. ***For a molecule with $z$ atoms, the
adjacency matrix will be $z \times z.$}
\begin{ruledtabular}
\begin{tabular}{llcc}
\textrm{Feature}& \textrm{Description} & \textrm{Type}& \textrm{Length}\\
\colrule
\textrm{Atom Type} & \textrm{H, C, N, O or F} & \textrm{One-hot} & 5\\
\textrm{Absorber} & \textrm{Is the atom type $=A$} & \textrm{1 or 0} & 1 \\
\textrm{Donor} & \textrm{Atom is an e$^-$ donor} & \textrm{1 or 0} & 1 \\
\textrm{Acceptor} & \textrm{Atom is an e$^-$ acceptor} & \textrm{1 or 0} & 1 \\
\textrm{Aromatic} & \textrm{Atom in aromatic ring} & \textrm{1 or 0} & 1 \\
\textrm{Hybridization} & \text{sp, sp$^2$ or sp$^3$} & \textrm{One-hot*} & 3 \\
\textrm{Bond Length} & \textrm{Distance in~\AA} & \textrm{Float} & 1 \\
\textrm{Bond Type} & \textrm{**} & \textrm{One-hot} & 4 \\
\textrm{Adjacency} & \textrm{Atomic connectivity} & \textrm{Matrix} & ***
\end{tabular}
\end{ruledtabular}
\end{table}

\section{Technical Details of the Message Passing Neural Network} \label{mpnn operation}
\begin{figure*}[!htb]
\includegraphics[width=\columnwidth]{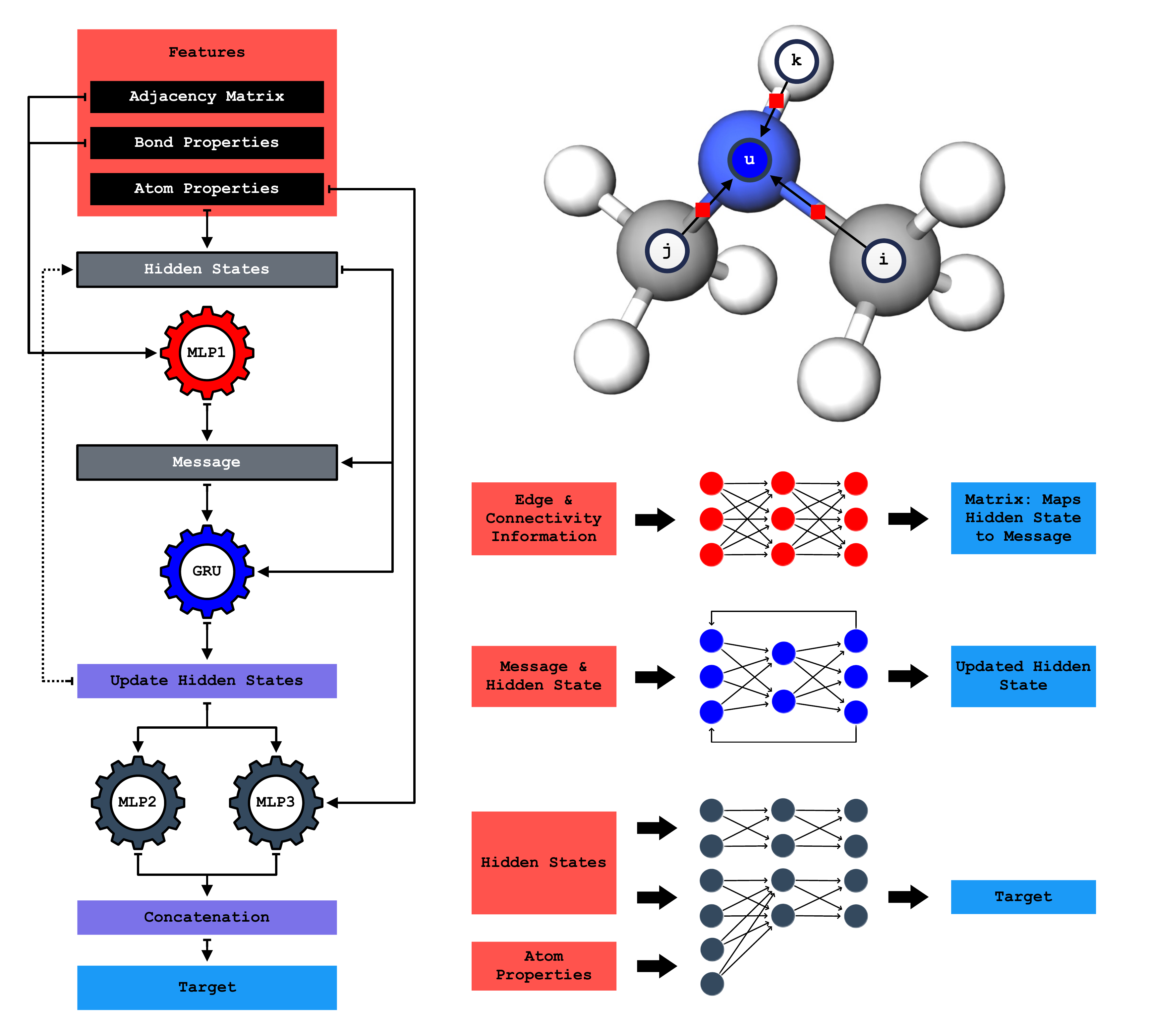}
\caption{\label{fig:mpnn_framework} Flowchart of the operation of the MPNN. Multi-layer perceptrons (MLPs) are denoted by gears, operations by
purple blocks, and internal hidden variables in grey. Initial features are contained in the larger red
square, and the target in light blue. Corresponding color-coded cartoons of the networks are shown to the
right of the flowchart (note these are not representative of the actual architectures). Finally, above the
cartoons is an example graph/molecule during the message passing (black arrows) and update (blue absorbing
nitrogen atom) phases. The red squares indicate that edge and nearest neighbor information
are used in MLP1 to construct the message.}
\end{figure*}

Here, we summarize the operation of the MPNN used in this work. To aid in visualization, we present
a flowchart of the MPNN operation in Fig.~\ref{fig:mpnn_framework}. 
For further details as well as
more examples of different MPNN architectures, we refer the reader to Ref. \cite{MPNN_theory}.
The general architecture of the MPNN is defined by two phases and three core functions. In the first
of the two phases, called the message passing phase (MPP), $n_i$ hidden
states, $h_v^0,$ are initialized for $v=1, ..., n_i,$ where $n_i$ is the number of atoms in molecule
$m_i,$ and $v$ represents an atom in that molecule, for every graph in $\mathcal{G}_A.$ For the
remainder of this section, we drop the molecular index and
assume features and targets refer to a single molecule. The initial hidden state vector is given by
\begin{equation}
    h_v^0 = (\mathbf{V}, 0, ..., 0), \quad \text{len}(h_v^0) = d.
\end{equation}
It means that the initial hidden state is simply the list of features for that atom, zero-padded up to
some length $d,$ which is a hyperparameter for the MPNN.

The MPP runs for $T$ iterations, and at each iteration, the hidden state vectors $h_v^t$ are reassigned via
the following procedure. First, a message
\begin{equation} \label{message function}
    m_v^{t+1} = \sum_{w\in N(v)}M^t(h_w^t,e_{vw})
\end{equation}
is computed by a sum of message functions $M^t.$ The message function incorporates information
about all nearest neighbor atoms $N(v),$ and all connecting bonds. Here, we follow the edge network
representation for $M^t$ as defined in \cite{MPNN_theory}, in which $M^t$ has the following properties:
$M^t = M,$ meaning that the \emph{same} message function is trained regardless of the time step $t$. 
$M$ is expressed as
\begin{equation}
    M(h_w^t, e_{vw}) = \text{MLP1}(e_{vw})h_w^t,
\end{equation}
where $e_{vw}$ represents the edge (bond) between nodes (atoms) $v$ and $w.$
The message function consists of a neural network trained on bond information only, i.e., for any single
atom, it is trained only on the bonds connecting it to its nearest neighbors. The output of MLP1 is a
$d\times d$ matrix. In this way, MLP1 learns a mapping between adjacent atoms $h_w$ and a
component of the message that ultimately ends up mutating the hidden state $h_v.$

For each $v,$ once the message has been computed at time step $t,$ the hidden states are updated via
the update function,
\begin{equation}\label{update function}
    h_v^{t+1} = U^t(h_v^t, m_v^{t+1}).
\end{equation}
The update function used in this work is a gated recurrent unit (GRU) introduced by Cho \emph{et al} \cite{gru}. It takes as input the current hidden state $h_v^t,$ as well as the
constructed message $m_v^{t+1}$ from the message function, and outputs the updated hidden
state $h_v^{t+1}.$ Importantly, like the message function, we choose $U^t = U$ such that
the GRU is independent of the time step. The independence of both the message
and update functions on the time step forces the respective networks to generalize to \emph{any} time step, and can
be thought of as a mechanism to guard against over fitting to the training data.

Finally, once the message passing phase has completed, the MPNN enters the readout phase,
which takes as input information about the hidden states (generally $h_v^0, h_v^1, ..., h_v^T$)
and predicts the target. The readout function
\begin{equation}
    \hat y = R(h_v^0, h_v^T) = \sum_v  \sigma\left[
    \text{MLP3}(h_v^0, h_v^T)\right] \circ \text{MLP2}(h_v^T)
\end{equation}
inputs the first and last hidden states (where the first hidden states $h_v^0$ are just the zero-padded
atom features) and outputs the length 80 target XAS spectrum. Note that $\circ$ denotes the element-wise
product, and $\sigma$ is the sigmoid activation function,
$$ \sigma(x) = [1 + e^{-x}]^{-1}.$$
While MLP2 is trained to make predictions based only on the last output of the MPP,
MLP3 effectively learns how to weight those predictions. The sigmoid activation function $\sigma(x)$
has range $(0, 1),$ and thus $\sigma\left[\text{MLP3}(h_v^0, h_v^T)\right]$ learns independently from
MLP2 how to weight the importance of the outputs, taking as input not only the last output of the MPP,
but also the initial atom features.

Training was conducted over 150 epochs using the Adam optimizer, and the Rectified Linear Unit
was used for all activation functions. All MLPs contained three hidden layers, each consisting
of 92 neurons, and the un-regularized $L_1$ loss was used as the cost function. No explicit
regularization (or dropout) was used during training, but effective regularization was manifest
through the use of identical message and update functions regardless of the time step.
The specific MPNN parameters, $T=3$ and $h=64,$ the maximum time step and length of the hidden
state vectors, respectively, were used throughout all calculations. All code was written in
PyTorch \cite{pytorch}, and training was conducted using a batch size of 512, distributed
evenly across parallel GPU architecture. 

\section{Learning Curves}
A hallmark of proper machine learning training is a systematic decrease in the loss metric with respect
to training set size until perceptive ability of the model is reached. Here, we show the loss functions for both the O and N data sets. We observe that all losses systematically decrease as the training size is increased.

\begin{figure*}[!htb]
\includegraphics[width=1.0\columnwidth]{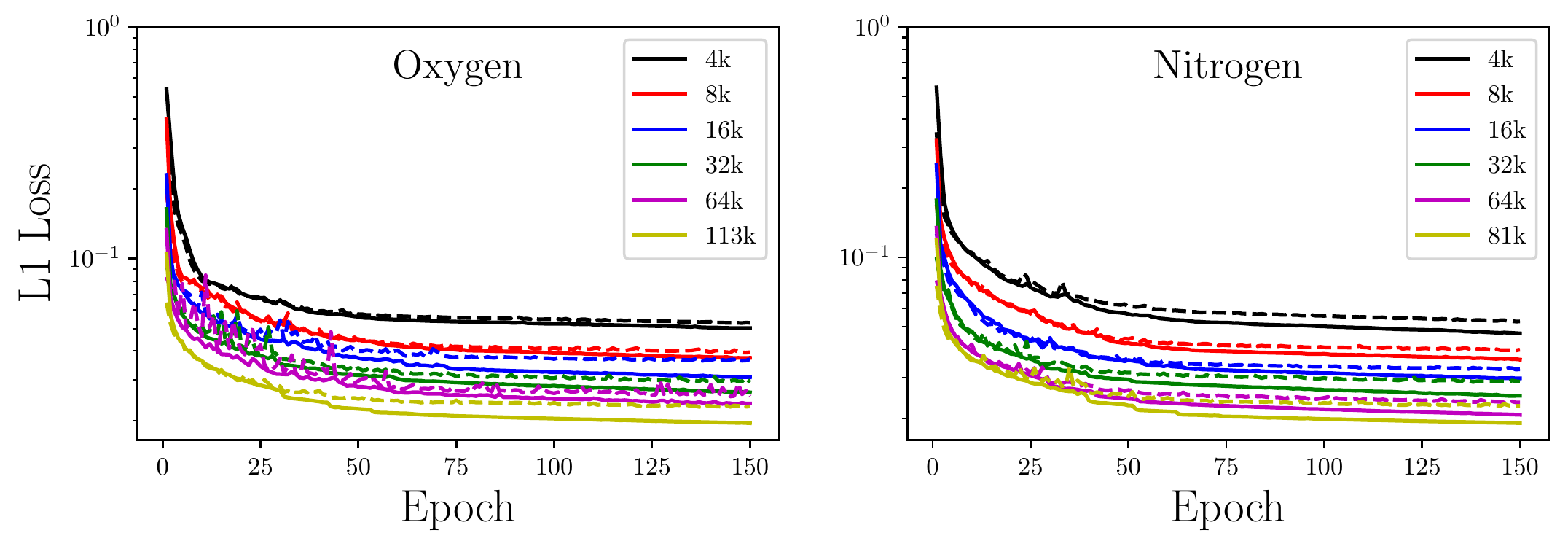}
\caption{\label{fig:losses} L1 loss of the oxygen and nitrogen data sets as a function of
training set size (see legend) and epoch. The largest training set size corresponds to the
full training set available for that database. Solid lines: training losses; dashed lines:
validation losses.}
\end{figure*}

\providecommand{\noopsort}[1]{}\providecommand{\singleletter}[1]{#1}%